\documentclass[]{aastex63}

\shorttitle{JWST/NIRCam imaging of NGC~6537}
\shortauthors{Kastner et al.}

\begin{document}

\title{JWST/NIRCam imaging of the Bipolar Planetary Nebula NGC~6537: the (Infra)red Spider, Revealed}

\correspondingauthor{Joel Kastner}
\email{joel.kastner@rit.edu}

\author[0000-0002-3138-8250]{Joel H. Kastner}
\affiliation{Center for Imaging Science, 
  Rochester Institute of Technology, Rochester NY 14623, USA; joel.kastner@rit.edu}
\affiliation{School of Physics and Astronomy, 
  Rochester Institute of Technology} 
\affiliation{Laboratory for Multiwavelength Astrophysics, 
  Rochester Institute of Technology}

\author{Paula Moraga Baez} 
\affiliation{School of Physics and Astronomy, 
  Rochester Institute of Technology} 
\affiliation{Laboratory for Multiwavelength Astrophysics, 
  Rochester Institute of Technology}

\author{Bruce Balick} 
\affiliation{Department of Astronomy, University of Washington, Seattle, WA 98195, USA} 
  
\author{Rodolfo Montez Jr.}
\affiliation{Center for Astrophysics $|$ Harvard \& Smithsonian, 60 Garden Street, Cambridge, MA 02138, USA}

\author{Caroline Gieser}
\affiliation{Max-Planck-Institut f\"{u}r Astronomie, K\"{o}nigstuhl 17, D-69117 Heidelberg, Germany}

\author{Mikako Matsuura}
\affiliation{Cardiff Hub for Astrophysics Research and Technology (CHART), School of Physics and Astronomy, Cardiff University,
The Parade, Cardiff CF24 3AA, UK}

\author{Jason Nordhaus}
\affiliation{School of Physics and Astronomy, 
  Rochester Institute of Technology} 
\affiliation{National Technical Institute for the Deaf, 
  Rochester Institute of Technology} 
  
\author{Miguel Santander-Garcia}
\affiliation{Observatorio Astronómico Nacional, C/ Alfonso XII 3\&5, E-28014 Madrid, Spain}

\begin{abstract}
We present {\it James Webb Space Telescope} (JWST) near-infrared (NIRCam) Br$\alpha$, H$_2$, [Fe {\sc ii}], and PAH imaging of the molecule-rich, high-excitation bipolar planetary nebula (PN) NGC 6537 (the Red Spider), complemented by new ALMA and Chandra observations and archival HST images. The resulting multiwavelength view of the Red Spider establishes the detailed lobe/torus structure of the nebula and the mass-loss history of its progenitor star. The extinction-penetrating JWST/NIRCam Br$\alpha$ and PAH and ALMA 3 mm continuum imaging exposes the complexity of the ionized inner nebula. JWST/NIRCam H$_2$ imaging traces the full, $\sim$1.1~pc extent of the bubble-like lobes formed by fast ($\sim$300--400 km s$^{-1}$) polar outflows, while ALMA $^{13}$CO(1--0) mapping reveals a point-symmetric, slowly ($\sim$10 km s$^{-1}$) expanding equatorial torus of radius $\sim$0.13~pc. In striking contrast, the [Fe~{\sc ii}] image displays an extended S-shaped emission morphology that traces
collisions between an active, collimated wind and slower-moving material along the lobe rims. No X-rays are detected from the nebula or its central star in deep Chandra/HRC-I imaging. However, the combined HST and JWST imaging reveals a near-IR excess at the central star indicative of emission from hot ($\sim$1000 K) circumstellar dust. We propose that interactions between the nebular progenitor star and a close companion are responsible for the ejection of NGC~6537's molecular torus, the formation of a circumbinary dust disk, and the launching of fast, wandering, collimated outflows that have inflated the polar lobe bubbles traced by near-IR H$_2$ emission and are presently generating the [Fe~{\sc ii}]-emitting shocks.

\end{abstract}

\section{Introduction}

The planetary nebula (PN) NGC 6537 is an archetype of the class of extreme high-excitation, yet highly molecule-rich, bipolar nebulae. This elite PN class includes both dynamically young objects such as NGC 7027, NGC 6302, Hubble 5, NGC 2440, NGC 6445, and dynamically older PNe, such as NGC~2818 and NGC~2899,  all of which are characterized by the presence of dusty, pinched waists and high-velocity polar outflows \citep[][and references therein]{Bublitz2023,Kastner2023,MoragaBaez2023,MoragaBaez2025}. Despite their large masses of molecular gas, these objects harbor some of the hottest and most luminous known PN central stars, with effective temperatures of $\sim$150--250~kK and luminosities of $\sim$$10^3$--$10^4$ K. The profound departures from spherical symmetry of such PNe defy the standard model of PNe as the products of single stars that have completed their asymptotic giant branch (AGB) evolution with the ejection of their H-rich envelopes, exposing their hot cores en route to becoming white dwarfs. Instead, the axisymmetric and point-symmetric density structures that characterize molecule-rich bipolar PNe are indicative of the presence of close, interacting binary companions \citep[e.g.,][and references therein]{DeMarco2009,JonesBoffin2017,Kastner2022}. 

Various models of the structural evolution of bipolar PNe, ranging from schematic to highly detailed, have been developed over the past four decades \citep[see, e.g.,][and references therein]{Morris1987,BalickFrank2002,Balick2019}. Most such models invoke either the formation of an accretion disk around the secondary star in a detached binary configuration with the mass-losing (PN-generating) AGB or post-AGB primary star \citep[e.g.,][]{Morris1987,Soker:1994ab,Soker:2000ab,Sahai2016,Chen2017,Huang2020}, or the onset of common envelope evolution in a closer initial binary configuration \citep[e.g.,][]{GarciaSegura2018,GarciaSegura2022,Zou2020}. However, direct evidence of binary companions to the central stars of high-excitation, molecule-rich bipolar PNe like NGC~6537 remains elusive, due to the bright, morphologically complex core regions of such PNe and the large  luminosities of their post-AGB (pre-white dwarf) central stars.

NGC 6537 has long been the subject of various ground-based optical imaging and spectroscopic studies aimed at understanding its structure, excitation, composition, and kinematics \citep[e.g.,][]{Feibelman1985,CorradiSchwarz1993,Cuesta1995,Aller1999}. In optical emission-line and near-IR H$_2$ imaging, the nebula displays a bright, compact core region with a pair of protruding, limb-brightened polar lobes, each extending over $\sim$1$'$ \citep{CorradiSchwarz1993,Kastner1996,Matsuura2005N6537}. 
This morphology, as captured in a color rendering of  {\it Hubble Space Telescope} (HST) optical imaging of the nebula,  earned NGC 6537 its ``Red Spider'' nickname\footnote{See https://apod.nasa.gov/apod/ap980106.html. This arthropodic ``Red Spider'' moniker has stuck (and is included in the nebula's {\tt simbad} database listing), despite NGC 6537's longstanding identification as a member of the class of pinched-waist ``butterfly'' nebulae \citep{CorradiSchwarz1993}.}. 
The distance to NGC 6537 is poorly determined, with literature estimates ranging from $D \sim 0.9$ to 2.4 kpc \citep{Matsuura2005N6537}; in this paper, we adopt the estimate $D=1.8\pm0.5$ kpc \citep{Frew2016}. 

The long-slit spectroscopy of \citet{CorradiSchwarz1993} and \citet{Cuesta1995} revealed (deprojected) expansion speeds of $\sim$300--400 km s$^{-1}$ along the nebula's polar axis, and constrained the polar axis inclination to lie in the range $\sim$30--40$^\circ$ with respect to the plane of the sky. \citet{Cuesta1995} also found spectroscopic evidence for expansion velocities at the nebular core that are both far higher ($\sim$2000 km s$^{-1}$) and much lower ($\sim$18 km s$^{-1}$) than those of the polar lobes, and modeled the ionized gas kinematics in terms of a disk/lobe structure. The notion of a system of polar lobe shocks in the model presented by \citet{Cuesta1995}  is supported by the nature of certain glaring deficiencies in photoionization modeling of the nebula \citep{Aller1999}, although the large ``disk'' in this model is perhaps better termed a ``torus'' given this large, dense, equatorial structure was not hypothesized to be in Keplerian rotation.

Many parallels between NGC 6537 and the intensively studied bipolar PN NGC 6302 have been described in the literature over the past four decades; indeed, the two have frequently been the subjects of observations reported and analyzed in the same paper. Both nebulae are often cited as classical examples of Peimbert Type I PNe, i.e., the N-rich descendants of relatively high-mass ($\sim$3--8 $M_\odot$) progenitor stars \citep[e.g.,][]{Quireza2007}. \citet{AshleyHyland1988} obtained detections of 1.96 $\mu$m [Si~{\sc vi}] emission in both PNe, implying very high central star effective temperatures in both cases. \citet{Rowlands1994} and \citet{Casassus2000} presented detections and analysis of numerous additional coronal (forbidden, high-ionization-state) IR lines from ISO and ground-based spectroscopy, respectively, of both nebulae.  \citet{Casassus2000} used their results to deduce central star effective temperatures of $\sim$240 kK for NGC 6302 and $\sim$160 kK for NGC 6537, and obtained luminosity estimates of 8500 $L_\odot$ for both stars.  On the basis of the detection of the central star of NGC 6537 in HST imaging, \citet{Matsuura2005N6537} obtained estimates for effective temperature and luminosity consistent with those inferred by \citet{Casassus2000}, albeit with much larger ranges of uncertainty in both parameters. 

The mm-wave molecular line surveys of NGC 6537 by \citet{EdwardsZiurys2013} and NGC 6302 by \citet{MoragaBaez2023,MoragaBaez2025} have established the extraordinarily rich molecular chemistries of both nebulae. In particular, both studies yielded numerous detections of N- and S-bearing molecules, indicative of the products of nucleosynthesis in the relatively massive progenitors of these PNe.  Interferometric mapping by ALMA demonstrates that, in both cases, the molecular gas is confined to dense, dusty equatorial tori expanding at $\sim$10--20 km~s$^{-1}$ \citep{MoragaBaez2023,MoragaBaez2025}.
On the basis of coherent submm polarization signatures, indicative of large-scale dust grain alignments, \citet{Sabin2007} found evidence for toroidal magnetic fields threading the dusty waists of both nebulae.

In this paper, we present a new observational study of NGC 6537 with the near-IR camera (NIRCam) aboard the {\it James Webb Space Telescope} (JWST), augmented by new observations with the {\it Chandra X-ray Observatory} (Chandra) as well as archival ALMA and HST data. The new JWST/NIRCam images of NGC 6537, combined with archival HST imaging, reveal the PN's detailed molecular and ionized gas and dust structures and the optical/near-IR spectral energy distribution of its central star, while the ALMA mm-wave mapping traces the ionized vs.\ neutral gas stratification and the kinematics of the dense regions in and around the bright core of NGC~6537. The Chandra observations are aimed at detecting X-rays from energetic shocks resulting from collisions between the fast ($\sim$300--400 km s$^{-1}$) collimated winds from the vicinity of the central star and slow ($\sim$10 km s$^{-1}$), dense, ``fossil'' AGB winds \citep[e.g.,][and references therein]{Kastner2012}, as well as to constrain the coronal activity of a potential companion to the central star  \citep[e.g.,][and references therein]{Montez2015}. We combine the results of these multiwavelength observations of NGC~6537, which span six decades of the electromagnetic spectrum, to elucidate the structure and formation of the Red Spider nebula and the nature of its progenitor star, so as to understand the origin and evolution of molecule-rich, high-excitation PNe more generally. 

The JWST and Chandra observations and archival data sources are described in \S 2. The JWST images and multiwavelength (JWST+ALMA+HST)  views of NGC 6537, and results derived from the imaging, are presented in \S 3. In \S 4, we discuss the insights into the structure and evolution of NGC 6537 obtained from the JWST+ALMA+HST imaging as well as hydrodynamic modeling of the formation of a bipolar PN, and we present analysis of the spectral energy distribution of the PN central star as yielded by the JWST+HST data. Section 5 presents a summary and conclusions.

\section{Observations}\label{sec:obs}

\subsection{JWST}

JWST/NIRCam images of NGC 6537 were obtained on 17 August 2024 under JWST PID 4571 (PI: J. Kastner), as part of a Joint Chandra/JWST program (see below). Images were obtained with NIRCam through filters F164N, F212N, F356W, and F405N, which isolate (respectively) [Fe {\sc ii}] 1.64 $\mu$m, H$_2$ S(1) 1--0  2.12 $\mu$m,  PAH 3.28 $\mu$m, and Br$\alpha$ 4.05 $\mu$m emission lines and features. 
The four images were obtained in pairs, using the SW+LW channels, via two dithered observations of NGC 6537. We employed SHALLOW2 readout with 6 groups per integration and 1 integration per exposure, and INTERMODULEX dithering with 8 primary dithers. These observing parameters yielded an exposure time of 2319 s per filter. 

Image processing tasks, as well as photometric and initial astrometric calibrations, were performed using standard JWST processing pipelines. Specifically, raw data products were generated from the original spacecraft telemetry using version {\tt 2024\_1a} of the JWST Science Data Processing subsystem. Version 1.14.0 of the JWST Science Calibration Pipeline was used to generate fully calibrated exposures and final mosaiced images for the dithered image sequences through each of the four filters. As the F356W dithered image sequence was saturated in the nebular core region after this standard pipeline processing, we re-ran the Science Calibration Pipeline using version 1.15.1 with the pixel ramp fitting option enabled during Level 1 processing (by setting the {\tt ramp\_fit} parameter {\tt suppress\_one\_group} to `False'). This additional processing step mitigated the saturation issues in the regions of the F356W image dominated by nebulosity, even though the core regions of field stars in the images remain saturated. 

Finally, the images output by the pipeline processing just described were astrometrically calibrated using the positions of field stars included in Gaia Data Release 2. While the relative (filter to filter) astrometric accuracy was excellent, we found position tweaks of order $\sim$0.5$''$ were necessary to bring the images into alignment with the Gaia star reference positions. We estimate the final, astrometrically calibrated images have absolute positional uncertainties of $\sim$0.05$''$.

\subsection{Chandra X-ray Observatory}\label{sec:Chandra}

We obtained six exposures targeting NGC 6537 with the High Resolution Camera imaging array (HRC-I) aboard the {\it Chandra X-ray Observatory} between 2024 Oct.\ and 2025 July (ObsIDs 28002, 28736, 28737, 28738, 28739, 31001), during Chandra Cycle 25, as part of Joint Chandra/JWST program 25200170 (PI: Kastner). The total Chandra/HRC-I exposure time was 76.13 ks. The resulting event data resulting from each exposure were processed through the Chandra X-ray Center's (CXC's) standard Chandra/HCR-I processing pipeline (version 10.13 for ObsIDs 28737 and 28738, and version 10.14.1.1 for all other ObsIDs), and each pipeline-processed event file was then subject to a hyperbolic screening algorithm to mitigate the effects of cosmic rays \citep{Kraft2018}. The six exposures were then merged. The resulting cosmic-ray-screened, merged (76.13 ks exposure) image yielded no detection of X-rays above the background within the area spanned by the NGC 6537 nebula in the JWST images (see \S \ref{sec:results1}). We discuss the significance of the X-ray nondetection of NGC 6537 in \S \ref{sec:XrayLimits}. 

\newpage

\subsection{Archival HST and ALMA data}

We have made use of archival data from HST and the {\it Atacama Large Millimeter Array} (ALMA) as points of comparison with our new JWST/NIRCam near-IR imaging of NGC 6537. As noted, the HST data, obtained in 1998 with WFPC2 (PID: 6502; PI: B. Balick), were included in the study of the core region of NGC 6537 by \citet[][see their Table 1]{Matsuura2005N6537}. The ALMA data were obtained in the 3 mm region (ALMA Band 3) as part of a molecular line survey of high-mass star-forming regions \citep[ALMA project code 2018.1.00424.S, PI: C. Gieser;][]{Gieser2023}, and are the subject of a forthcoming paper (Moraga Baez et al.\ 2025b, in preparation). Here, we utilize maps of 98.3 GHz continuum emission and the $J=1\rightarrow 0$ (110.20135 GHz) rotational transition of $^{13}$CO, obtained at  spatial resolution of $\sim$1$''$ \citep[see][for details]{Gieser2023}. The $^{13}$CO emission-line map was obtained with velocity resolution of 0.664 km s$^{-1}$, providing a data cube from which we have extracted velocity-integrated (``moment 0'') images over various velocity ranges of interest as well as an intensity-weighted mean velocity (``moment 1'') image.

\section{Results}

\subsection{Individual JWST/NIRCam images: full-field and zoomed views}\label{sec:results1}

\begin{figure}[ht]
\begin{center}
\includegraphics[width=5in]{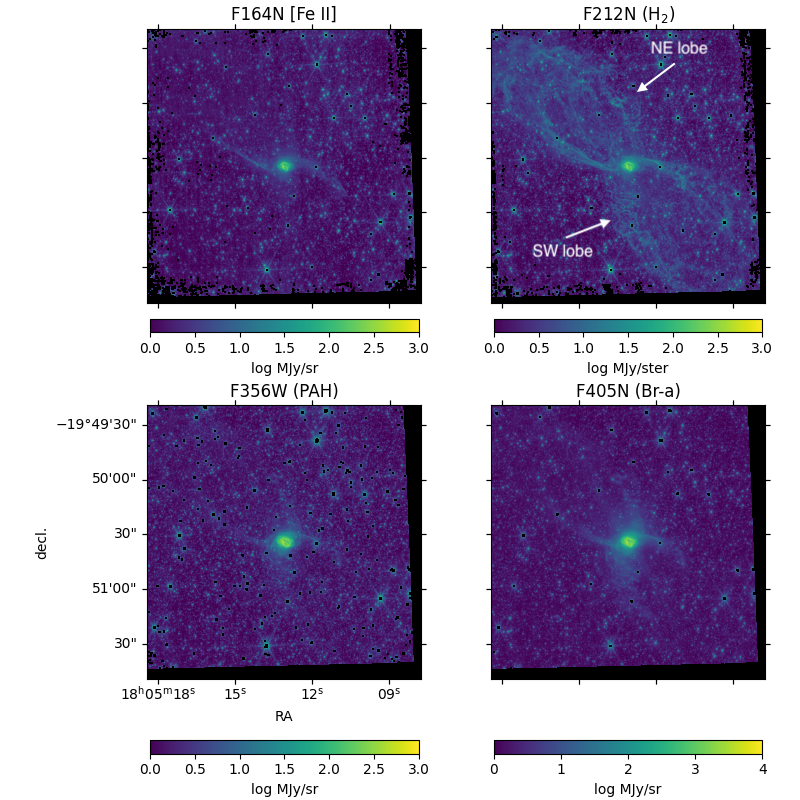}
\end{center}
\caption{JWST/NIRCam images of NGC~6537, full $150''\times150''$ field of view (FOV). The four image frames are arranged, from left to right and top to bottom, in the order F164N, F212N, F356W, F405N. The rearward-facing NE and forward-facing SW polar lobes are indicated in the H$_2$ (F212N, upper right) image.}
\label{fig:JWSTfull}
\end{figure}

\begin{figure}[ht]
\begin{center}
\includegraphics[width=5in]{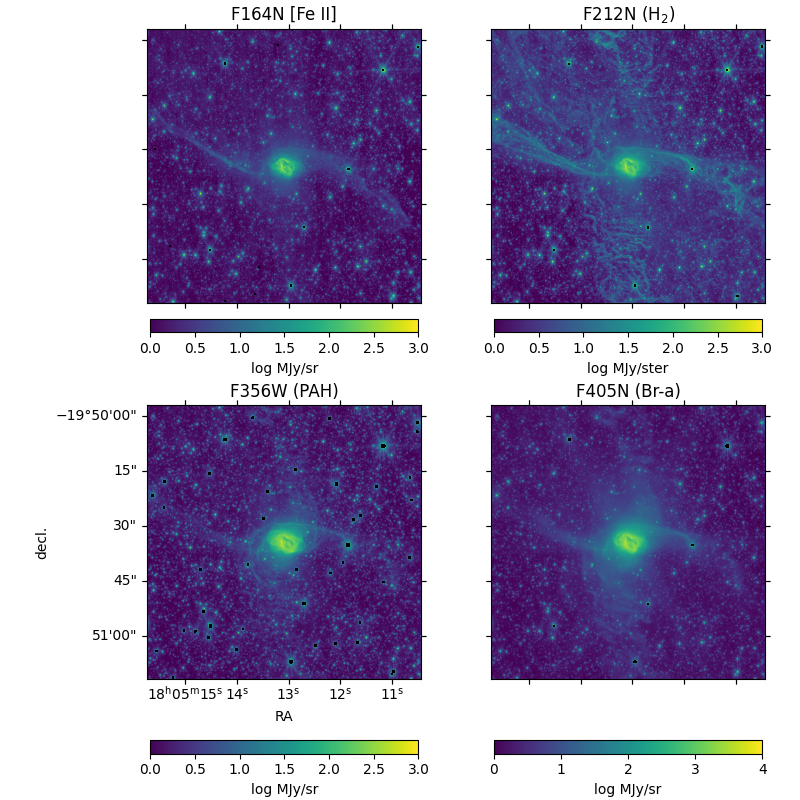}
\end{center}
\caption{JWST/NIRCam images of NGC~6537 arranged as in Fig.~\ref{fig:JWSTfull}; zoomed-in views, $75''\times75''$ FOV.}
\label{fig:JWSTzoom1}
\end{figure}

\begin{figure}[ht]
\begin{center}
\includegraphics[width=5in]{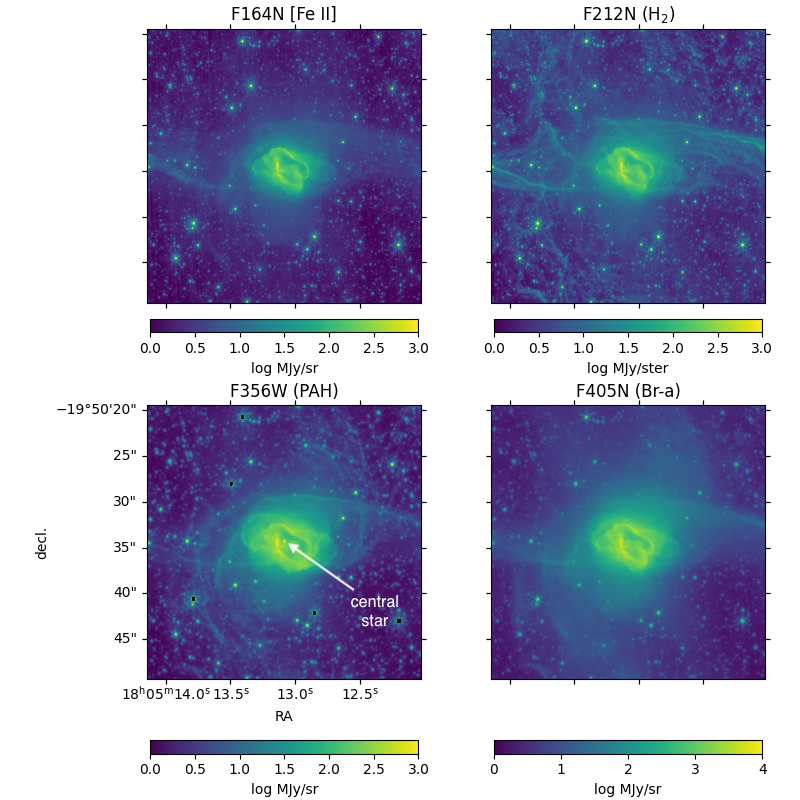}
\end{center}
\caption{JWST/NIRCam images of NGC~6537 arranged as in Fig.~\ref{fig:JWSTfull};  zoomed-in views, $30''\times30''$ FOV. The position of the central star is indicated in the  PAH (F356W, lower left) image.}
\label{fig:JWSTzoom2}
\end{figure}

The resulting full-field ($\sim$$150''\times150''$) JWST/NIRCam [Fe {\sc ii}], H$_2$, PAH, and Br$\alpha$ images of NGC 6537 are displayed in Figure~\ref{fig:JWSTfull}. Zoomed-in views of the JWST/NIRCam images, with fields of view of $\sim$$75''\times75''$ and $\sim$$30''\times30''$, are presented in Figure~\ref{fig:JWSTzoom1} and Figure~\ref{fig:JWSTzoom2}, respectively. The morphology of the 4.05 $\mu$m Br$\alpha$ image in the full-field rendering in Figure~\ref{fig:JWSTfull} --- with its bright core and open, limb-brightened polar lobes extending to the northeast and southwest from the core along position angle (PA) $\sim$45$^\circ$, forming opposed pairs of ``horns'' --- bears close resemblance to that seen in ground-based H$\alpha$ imaging of the nebula \citep[e.g.,][their Fig.~2]{Cuesta1995}. The ground-based spectroscopy has established that the northeast-directed and southwest-directed lobes (hereafter referred to as the NE and SW lobes) are directed away from and toward the observer, respectively \citep{CorradiSchwarz1993,Cuesta1995}. 

The most immediately striking aspects of the full-field JWST/NIRCam images in Figure~\ref{fig:JWSTfull} are the detection of bubble-like polar lobe enclosures in 2.12 $\mu$m H$_2$ emission, and the S-shaped  morphology of the 1.64~$\mu$m [Fe~{\sc ii}] emission. The bubble-like structures imaged in H$_2$, each of which extend $\sim$100$''$ from the nebular core region to the extreme corners of the NIRCam field of view in the H$_2$ image, were only marginally detected in previous ground-based 2.12 $\mu$m H$_2$ imaging \citep{Kastner1996,Marquez-Lugo2013}, and do not appear as complete bubble-like structures in previous, deep ground-based and HST H$\alpha$+[N~{\sc ii}] imaging \citep{CorradiSchwarz1993,Matsuura2005N6537,Marquez-Lugo2013}. For our adopted distance, $D=1.8$ kpc, the projected linear extent of each polar lobe as revealed in the JWST/NIRCam H$_2$ imaging is $\sim$0.9 pc. The deprojected lengths are then $\sim$1.1 pc \citep[assuming $i=38^\circ$;][]{CorradiSchwarz1993}. This implies lobe dynamical ages of $\sim$3700 yr, given the inferred (deprojected) lobe expansion speeds of $\sim$300 km s$^{-1}$  \citep[][]{CorradiSchwarz1993}. 

The zoomed-in ($\sim$$75''\times75''$) views in Figure~\ref{fig:JWSTzoom1} provide closer looks at the S-shaped [Fe {\sc ii}] emission morphology of the nebula and the delicate fine structures and wave-like features detected in H$_2$ emission. The profound point-symmetry of the [Fe {\sc ii}] emission morphology, which is also seen in lower-resolution (ground-based) 1.64 $\mu$m imaging of NGC 6537 \citep[][]{Kim2024}, is remarkably similar to that of NGC 6302, as first revealed by HST/WFC3 1.64 $\mu$m imaging \citep{Kastner2022}. 
The $\sim$$75''\times75''$ frames in Figure~\ref{fig:JWSTzoom1} also make evident the more compact appearance of the nebula in the F356W image relative to the images in the other three filters.
It is apparent from the $\sim$$30''\times30''$ image frames in Figure~\ref{fig:JWSTzoom2} that the fine structures seen in the F356W image just outside the bright nebular core region are similar to those imaged in H$_2$ emission. These zoomed-in views of the core region also demonstrate that NGC 6537's central star is well-detected in all four NIRCam images. The $\sim$$30''\times30''$ image frames furthermore highlight the complex structure of the nebular core region, as was also apparent in HST emission-line imaging \citep{Matsuura2005N6537}.

\subsection{JWST/NIRCam images: color image overlays}\label{sec:results2}

\begin{figure}[ht]
\begin{center}
\includegraphics[width=6in]{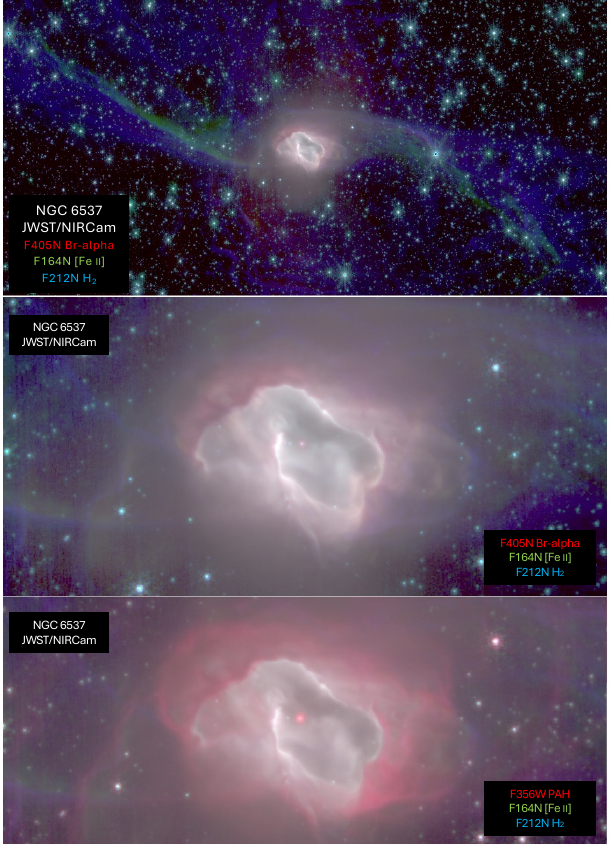}
\end{center}
\caption{Top and middle panels: Wide-field ($\sim$$80''\times40''$) and zoomed-in ($\sim$$20''\times10''$) color overlays of 4.05 $\mu$m Br$\alpha$ (red), 1.64 $\mu$m [Fe {\sc ii}] (green), and 2.12 $\mu$m H$_2$ (blue) images of NGC~6537. Bottom: the same ($\sim$$20''\times10''$) field of view as the middle panel, with the 3.56 $\mu$m PAH image replacing Br$\alpha$ in the red channel. Images are displayed in log scale, with min/max intensities as in Fig.~\ref{fig:JWSTfull}, with further enhancements to highlight emission outside the core regions. }
\label{fig:JWST_Bra_PAH_FeII_H2_RGB}
\end{figure}

Various key features of the structure of NGC 6537 are further highlighted by the color overlays of the NIRCam images presented in Figure~\ref{fig:JWST_Bra_PAH_FeII_H2_RGB}. The zoomed-in views  in the middle and bottom panels make clear the apparently multi-lobed yet profoundly asymmetric structures within the nebular core region. The white color of the sharp-rimmed filamentary structures threading this core region in Figure~\ref{fig:JWST_Bra_PAH_FeII_H2_RGB}, best seen in the middle and bottom panels, reflects the similar intensities in the three images included in each overlay, and is hence indicative of a significant contribution of continuum (free-free) emission within the F164N, F212N, and F356W images, in the nebular core. Ground-based imaging had previously established the core region as a bright near-IR continuum source \citep{Kastner1996}, but with insufficient resolution to elucidate the fine structures threading this region. Indeed, the bright rim of near-IR continuum emission, as revealed in the zoomed-in views in Figure~\ref{fig:JWST_Bra_PAH_FeII_H2_RGB} (middle and bottom panels), appears to wrap around the central star in an irregular spiral pattern.

Note that the central star appears red in both zoomed-in color overlays in Figure~\ref{fig:JWST_Bra_PAH_FeII_H2_RGB}. Given the central star's very high temperature \citep[$\sim$160 kK;][]{Casassus2000}, this red color is indicative of the presence of a near-IR excess at the star that is well detected in (at least) the F356W and F405N filters (both of which are coded red in the lower panels of Figure~\ref{fig:JWST_Bra_PAH_FeII_H2_RGB}). We analyze this excess quantitatively in \S~\ref{sec:IRexcess}.

Other differences between the NIRCam images that are highlighted in red, green, or blue in Figure~\ref{fig:JWST_Bra_PAH_FeII_H2_RGB} reveal emission in the specific line or feature lying within the filter bandpass. Thus the top panel of Figure~\ref{fig:JWST_Bra_PAH_FeII_H2_RGB} illustrates how, within both lobes, the S-shaped [Fe {\sc ii}] emitting region (color-coded green) lies adjacent to, but offset from, the H$_2$-emitting lobe rims (color-coded blue); more specifically, the [Fe {\sc ii}] is seen to trace the inner edges of the southern rim of the NE lobe and the northern rim of the SW lobe. Meanwhile comparison of the zoomed-in views of the core region in the middle and bottom panels of Figure~\ref{fig:JWST_Bra_PAH_FeII_H2_RGB} reveals a cocoon of luminous PAH emission (color-coded red) surrounding the ionized nebular core \citep[compare with Fig.~8 in][]{Matsuura2005N6537}. Just outside this core region, as previously noted, the PAH emission is observed to closely trace the fine filamentary structures detected in H$_2$; however, the surface brightness of PAH emission drops precipitously beyond the core and inner lobe regions, while the filamentary H$_2$ emission outlines the full extents of the polar lobes contained within the NIRCam field of view. The vast reach of the H$_2$ relative to the PAH emission is indicative of shock excitation for the former, vs.\ UV excitation for the latter.

\subsection{The JWST/HST/ALMA/Chandra multiwavelength view of the Red Spider}\label{sec:results3}

\subsubsection{The ionized nebular core}\label{sec:core}

\begin{figure}[ht]
\begin{center}
\includegraphics[width=6in]{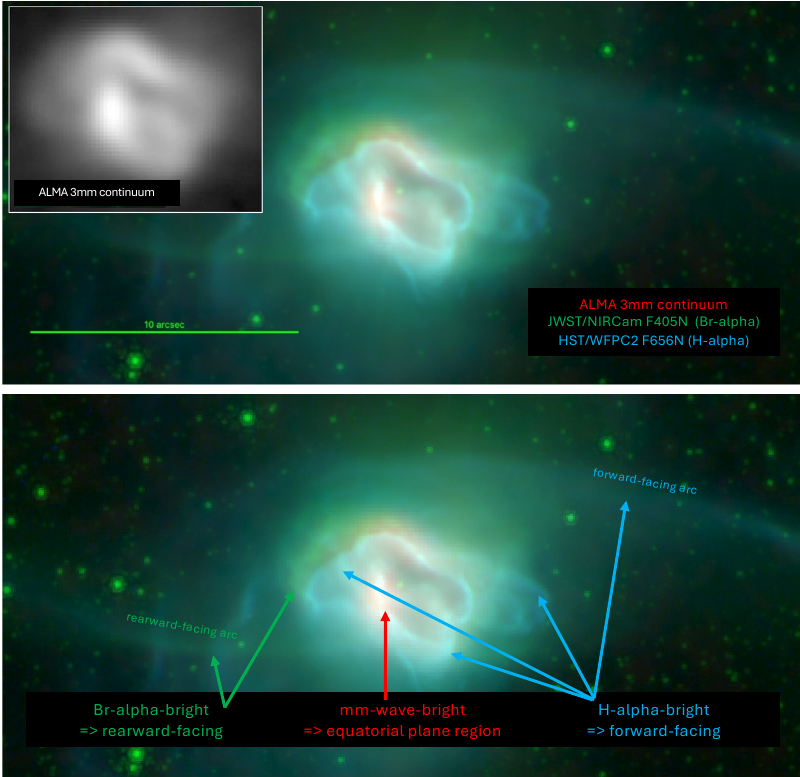}
\end{center}
\caption{Overlay of ALMA 3 mm continuum emission map (red), JWST/NIRCam 4.05 $\mu$m Br$\alpha$ image (green), and HST/WFPC2 H$\alpha$ image (blue) of NGC~6537. The ALMA continuum image is also presented in the B\&W inset of the top panel. The annotations in the bottom panel denote features that are especially bright in H$\alpha$, Br$\alpha$, or 3 mm continuum. Features that are bright in H$\alpha$ or Br$\alpha$ and are likely directed toward and away from the observer, respectively, while the mm-wave continuum likely traces ionized material confined to the PN's equatorial plane region (see \S \ref{sec:core}).}
\label{fig:3mm_Bra_Ha_RGB}
\end{figure}

Fig.~\ref{fig:3mm_Bra_Ha_RGB} presents a color overlay of the ALMA 3 mm continuum emission map (red) on the JWST/NIRCam 4.05 $\mu$m Br$\alpha$ (green) and HST/WFPC2 H$\alpha$ (blue) images for the core region of NGC~6537, with the 3 mm continuum map also included (in B\&W) as an inset in the top panel of the figure. Annotations in the bottom panel of Fig.~\ref{fig:3mm_Bra_Ha_RGB} point out features that are especially bright in H$\alpha$, Br$\alpha$, or 3 mm continuum. The H$\alpha$ emission, being the most susceptible to intranebular extinction, is most likely brightest within the regions of ionized gas that lie in the nebular foreground and are directed toward the observer (cyan-shaded regions in Fig.~\ref{fig:3mm_Bra_Ha_RGB}); conversely, the Br$\alpha$ emission likely highlights those ionized regions of the nebula that lie in the background and are directed away (green regions in Fig.~\ref{fig:3mm_Bra_Ha_RGB}). It is hence readily apparent that the three small ($\sim$3--5$''$ long) lobes or loops of emission near but outside the core that are directed to the E, SSW, and W of the central star are foreground/forward-directed features, as is the larger arc of emission extending from the E of the star around to its N and then to its WNW. The Br$\alpha$-bright diffuse region extending beyond and to the N of the forward-directed E loop, meanwhile, is directed away, as is the arc feature that wraps around to the S of the star and then to its ESE. The two arc features in Fig.~\ref{fig:3mm_Bra_Ha_RGB} are evidently the limb-brightened innermost edges of the oppositely directed polar lobes that are imaged in their entirety (in H$_2$ emission) in Figure~\ref{fig:JWSTfull}. 

Fig.~\ref{fig:3mm_Bra_Ha_RGB} further demonstrates that the 3 mm continuum emission, which is dominated by free-free emission from the ionized gas, traces the central, $\sim$2$''$ radius loop feature within the core (red regions in Fig.~\ref{fig:3mm_Bra_Ha_RGB}). The mm-wave emission from this dense plasma is not susceptible to dust obscuration, and so likely traces ionized gas that is confined to the equatorial plane of the system --- a zone sandwiched between, and hence seen projected within, the regions that appear bright in H$\alpha$ and Br$\alpha$. This suggests that the 3 mm continuum likely emanates from the ionized innermost regions of the dense, dusty equatorial torus whose outer regions are traced by mm-wave emission from $^{13}$CO (see next).

\subsubsection{The molecular torus}\label{sec:COtorus}

\begin{figure}[ht]
\begin{center}
\includegraphics[width=6.5in]{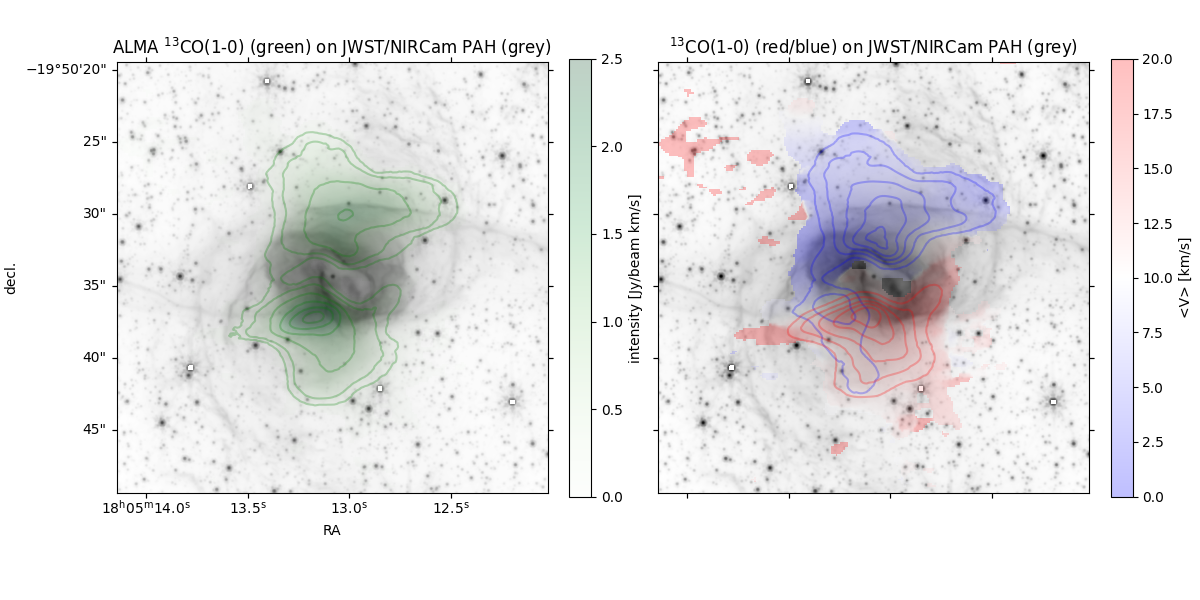}
\end{center}
\caption{{\it Left:} Overlay of velocity-integrated ALMA $^{13}$CO(1--0) emission-line map (green linear intensity scale, with green contours) on JWST/NIRCam 3.56 $\mu$m PAH (logarithmic greyscale). The contours are set to levels of 0.25, 0.5, 1.0, 1.5, 2.0, and 2.5 Jy beam$^{-1}$ km s$^{-1}$. {\it Right:} Intensity-weighted velocity map of ALMA $^{13}$CO(1--0) emission, showing blueshifted/redshifted emission (see color bar) on JWST/NIRCam 3.56 $\mu$m PAH (greyscale), with contours of velocity-integrated blueshifted (0--9 km s$^{-1}$) and redshifted (11--20 km s$^{-1}$) $^{13}$CO(1--0) emission overlaid; the contours are set to levels of 0.25, 0.5, 0.75, 1.0, 1.25, and 1.5 Jy beam$^{-1}$ km s$^{-1}$.}
\label{fig:ALMAvsJWST}
\end{figure}

Overlays of the ALMA $^{13}$CO(1--0) image data on the JWST/NIRCam 3.56 $\mu$m PAH+continuum (F356W) image of NGC~6537  are presented in Fig.~\ref{fig:ALMAvsJWST}. The left panel illustrates the relationship between velocity-integrated mm-wave $^{13}$CO(1--0) emission and the near-IR PAH and continuum emission. The overlay demonstrates that the cold, dense molecular gas traced in $^{13}$CO is confined to the equatorial regions of the nebula.
The $^{13}$CO emission in NGC 6537 is also distinctly point-symmetric, with bright emission NNW and SSE of the central star (along position angle PA $\sim$155$^\circ$) and a conspicuous lack of $^{13}$CO emission to its east and west (along position angle PA $\sim$90$^\circ$). The superposition of $^{13}$CO and PAH+continuum emission also reveals a central cavity in the molecular gas that is spatially coincident with the bright emission from ionized gas at the nebular core (i.e., the emission that dominates Fig.~\ref{fig:3mm_Bra_Ha_RGB}).

The right panel of Fig.~\ref{fig:ALMAvsJWST} reveals the radial velocity field of the $^{13}$CO(1--0) emission, demonstrating the expansion of the molecular gas from the vicinity of the central star. The data cube and moment 1 (intensity-weighted velocity) image are characterized by a projected expansion velocity of $V_{exp} \sim 10$ km s$^{-1}$ with respect to NGC 6537's systemic velocity of $\sim$10 km s$^{-1}$ \citep[][]{EdwardsZiurys2013}. This $V_{exp}$ is similar to that measured for the low-velocity component of ionized gas in the nebular core \citep[18 km s$^{-1}$;][]{Cuesta1995}, suggesting the molecular gas and the low-velocity ionized gas are two components of the same physical structure. Surprisingly, however, the blueshifted $^{13}$CO(1--0) line emission is seen superimposed on the western quadrant of the {\it rearward-facing} NE lobe and the redshifted line emission is projected against the eastern quadrant of the {\it forward-facing} SW lobe. This unexpected juxtaposition is in fact readily explained if we hypothesize that --- as in the case for other, analogous molecular-rich bipolar nebulae (e.g., NGC 6302; Moraga Baez et al.\ 2025) --- the molecular gas is confined to an expanding equatorial torus that is oriented nearly perpendicularly to the main outflow lobes. In the case of NGC 6537, it appears this expanding molecular torus is characterized by a profoundly point-symmetric density structure whose main direction of symmetry, like that of the 1.64 $\mu$m [Fe {\sc ii}] emission, is significantly out of alignment with the polar axis of the nebula as seen projected against the plane of the sky. Specifically, given the polar lobe axis of symmetry (PA $\sim$45$^\circ$; Figure~\ref{fig:JWSTfull}), the projected orientation of peak $^{13}$CO emission (PA $\sim$155$^\circ$) is evidently rotated by $\sim$20$^\circ$ with respect to the expected (projected) orientation of the nebula's equatorial plane (i.e., PA $\sim$135$^\circ$). 

Notwithstanding its point-symmetry and misalignment, if we assume the inclination of the molecular torus axis of symmetry with respect to the plane of the sky is identical that of NGC 6537's polar axis \citep[38$^\circ$;][]{CorradiSchwarz1993}, then the deprojected expansion velocity of the torus we obtain from the $^{13}$CO(1--0) data is $\sim$13 km s$^{-1}$.  From Fig.~\ref{fig:ALMAvsJWST}, we deduce a projected angular radius of $\sim$9$''$ for the outer edge of the molecular torus, leading to a deprojected linear radius of $\sim$0.13 pc assuming $D=1.8$ kpc. The implied torus dynamical age --- or, more accurately, the time since onset of copious equatorial mass loss --- is then $\sim$10000 yr. This timescale is significantly larger than the $\sim$3700 yr dynamical age of the lobes, given their extents in the JWST H$_2$ imaging and their deprojected velocities as inferred from ground-based spectroscopy (\S~\ref{sec:results1}). Such a disparity in lobe vs.\ torus ages --- wherein the relatively slow outflow of dense material along the equatorial plane appears to have been initiated well prior to the formation of the fast outflows that generated the rarified polar lobes --- appears to be a defining characteristic of molecule-rich, bipolar PNe \citep{MoragaBaez2025} and PNe with torus/lobe morphologies more generally \citep[e.g.,][]{Huggins2007,GarciaDiaz2009}. We further consider this aspect of the structural evolution of NGC 6537 in \S~\ref{sec:modeling}.

\subsubsection{Upper limits on X-ray emission from NGC 6537}\label{sec:XrayLimits}

The Chandra planetary nebula survey (ChanPlaNS) demonstrated that $\sim$30--40\% of PNe exhibit either diffuse (extended), soft X-ray emission from rarified, shock-heated gas (``hot bubbles'') generated by wind collisions \citep[][and references therein]{Kastner2012,Freeman:2014fq} or a point-like and (typically) somewhat harder X-ray source likely associated with the corona of a putative companion star \citep[][and references therein]{Montez2015}. A subset of these X-ray-detected PNe exhibit both types of X-ray source. Our nondetection of X-rays in deep Chandra/HRC-I imaging of NGC 6537 (\S \ref{sec:Chandra}) hence provides constraints on both forms of PN high-energy phenomena associated within this nebula. 

To obtain upper limits on the X-ray luminosities of diffuse ($L_{X,d}$) and point-like ($L_{X,p}$) sources within NGC~6537, respectively, we measured the background counts in the  cosmic-ray-screened, 76.13 ks exposure Chandra/HRC-I image within a 40$''$ radius circular aperture centered on the nebula. Within this region, we recover 3131 events, for a background photon count rate of $8.12\times10^{-3}$ counts ks$^{-1}$ arcsec$^{-2}$. Applying this result to $3''\times2''$ elliptical ($\sim$18 arcsec$^2$) and $1.0''$ radius circular ($\sim$3 arcsec$^2$) regions encompassing the nebular core and central star, respectively (see \S \ref{sec:results2}), yields respective background count rates of 0.15$\pm$0.05 ks$^{-1}$ and 0.02$\pm$0.01 ks$^{-1}$, where the uncertainties reflect Poisson photon counting. The 5$\sigma$ upper limits on source count rates within the ionized core and central star regions implied by our nondetection of X-ray photons from NGC 6537 within these regions are thus $\sim$0.25 ks$^{-1}$ and $\sim$0.05 ks$^{-1}$, respectively. 

To convert these count rate upper limits to X-ray flux upper limits, we used the webPIMMS\footnote{https://heasarc.gsfc.nasa.gov/cgi-bin/Tools/w3pimms/w3pimms.pl} spectral modeling tool. We adopt the APEC thermal plasma model, assuming plasma temperatures of 3 MK and 10 MK for ionized core and central star source regions, respectively. The former is typical of the energetic shocks from collisions between the fast (present-day) central star and ambient (``fossil'' AGB) winds \citep[][and references therein]{Kastner2012,MontezKastner2018}, while the latter is typical of the coronae of active central star companions \citep[][and references therein]{Montez2015}. We further assume an intervening absorbing column of $N_H = 7\times10^{21}$ cm$^{-2}$, corresponding to the extinction toward the central star and nebular core region, $A_V=3.7$ \citep{Matsuura2005N6537}. 
For these choices of model parameters, the measured Chandra/HRC-I count rate limits correspond to webPIMMS-inferred upper limits on the intrinsic 0.2--10 keV fluxes of (diffuse) nebular core and (point-like) central star X-ray emission of $\sim$$4\times10^{-14}$ erg cm$^{-2}$ s$^{-1}$ and $\sim$$5\times10^{-15}$ erg cm$^{-2}$ s$^{-1}$, respectively. These flux limits yield limiting luminosities of $L_{X,d} \lesssim 1.5\times10^{31}$ erg s$^{-1}$ and $L_{X,p} \lesssim 2\times10^{30}$ erg s$^{-1}$ for the diffuse and point-like X-ray emission sources, respectively, given our assumed distance of 1.8 kpc. 

Putting these results into perspective, the upper limit on $L_{X,d}$ is a factor $\sim$5 smaller than the X-ray luminosities of the diffuse X-ray sources within the much younger (dynamical age $\sim$1000 yr), molecule-rich PNe NGC 7027 \citep{,MontezKastner2018} and BD+30$^\circ$ 3639 \citep{Kastner:2000rr}. Our nondetection of X-rays from energetic shocks in NGC 6537 is nevertheless consistent with the results of ChanPlaNS, which yielded few such detections of diffuse X-ray emission within molecule-rich PNe \citep{Kastner2012,Freeman:2014fq}. 
Meanwhile, the upper limit on $L_{X,p}$ appears to rule out the presence of a close, magnetically active, main-sequence companion star of bolometric luminosity $L_{bol} \gtrsim 2\times10^{33}$ erg s$^{-1}$, corresponding to G or earlier spectral type, that has been spun-up to coronal saturation levels (i.e., exhibits $\log{(L_{X,p}/L_{bol})} \sim -3$) via accretion of AGB ejecta  \citep[][and references therein]{Montez2015}. We stress, however, that the limits on both $L_{X,d}$ and $L_{X,p}$ derived here are dependent on the aforementioned choices of parameters (especially plasma temperatures) input to the respective X-ray source models.

\section{The structure and evolution of the Red Spider Nebula}\label{sec:modeling}

\subsection{Main structural components}\label{sec:schematic}

\begin{figure}[ht]
\begin{center}
\includegraphics[width=6.5in]{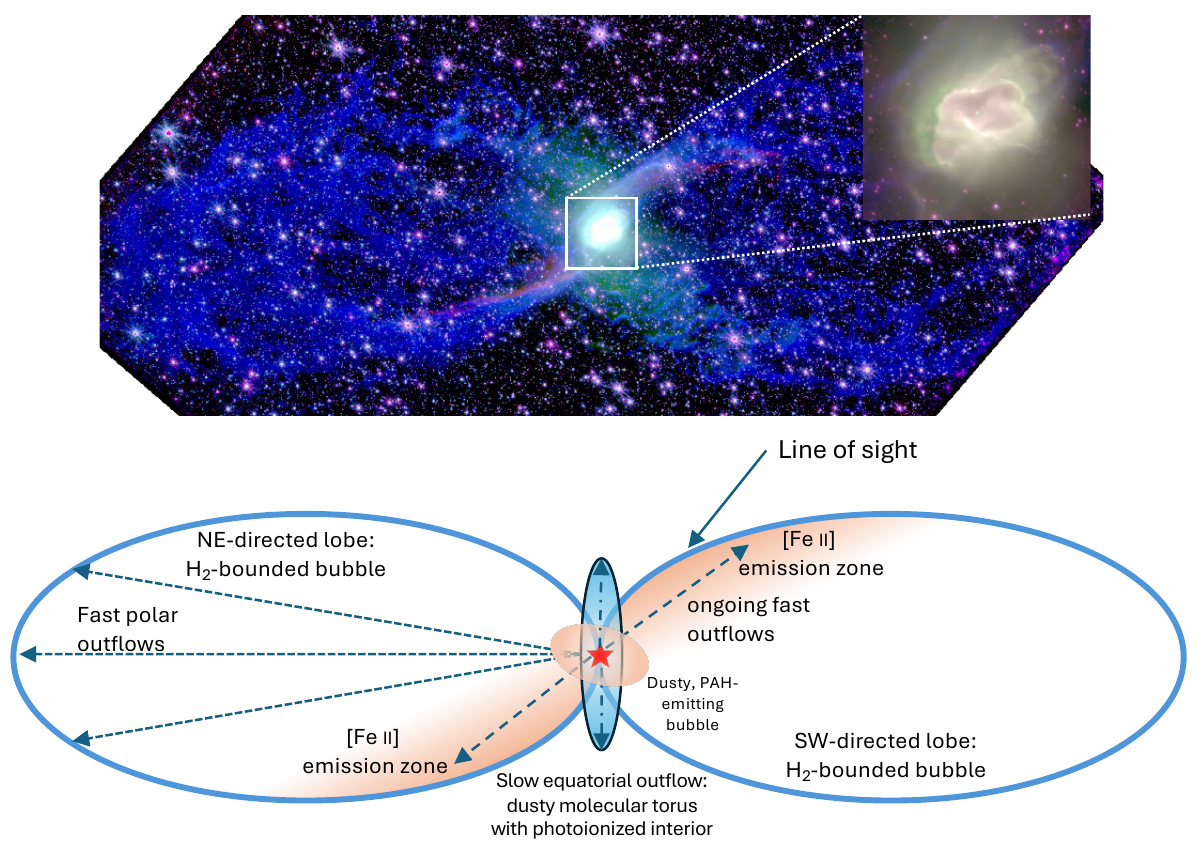}
\end{center}
\caption{Top: Overlay of 4.05 $\mu$m Br$\alpha$ (green), 2.12 $\mu$m H$_2$ (blue), and 1.64 $\mu$m [Fe {\sc ii}] (red) images of NGC~6537, rotated through 47$^\circ$ in PA (such that the polar axis is horizontal in the image). The inset shows a zoom into the core region, which is saturated in the wide-field rendering. Bottom: Schematic of the main structural components of NGC 6537: (1) H$_2$-bounded polar lobes; (2) point-symmetric, [Fe {\sc ii}]-emitting regions along opposing lobe walls; (3) slowly expanding equatorial molecular torus with ionized interior; (4) bubble-like dust structure detected in PAH emission. The line of sight toward the system (indicated at the upper right of the diagram) is such that both the polar axis and the equatorial torus are viewed at intermediate inclination. See text (\S~\ref{sec:schematic}). }
\label{fig:StructuralSchematic}
\end{figure}

In light of the results just presented, we have constructed a schematic view of the main structural components of NGC 6537. This schematic is presented in Figure~\ref{fig:StructuralSchematic}, below a reference color overlay of the JWST/NIRCam H$_2$, [Fe~{\sc ii}], and Br$\alpha$ images. The basic structural components, moving inwards, consist of: 
\begin{enumerate}
\item A pair of H$_2$-bounded polar lobes, with photoionized and/or shock-ionized inner rims, generated by the fast ($\sim$300-400 km s$^{-1}$) polar flows detected in ground-based long-slit spectroscopy \citep{CorradiSchwarz1993}. 
\item The point-symmetric, [Fe {\sc ii}]-emitting regions along opposing lobe walls, which likely mark the interaction zones where ongoing fast outflows are shearing and shocking the relatively dense, swept-up material along the lobe rims. 
\item A dense and dusty equatorial torus that is slowly expanding (expansion velocity $\sim$10 km s$^{-1}$). The torus is primarily composed of cold ($\lesssim$100 K), CO-emitting molecular gas, but its innermost region is heated and ionized by UV and (perhaps) fast winds from the central star, and is hence a luminous source of H recombination line emission and near-IR and 3 mm continuum (free-free) emission.
\item A bubble-like dust structure that appears as a zone of bright PAH emission surrounding the ionized core region. 
\end{enumerate}
The line of sight toward this system, indicated at the upper right of the schematic diagram in Figure~\ref{fig:StructuralSchematic}, is such that both the polar axis and the equatorial torus are viewed at intermediate inclination, so as to be consistent with the inferred polar axis inclination \citep[$\sim$30--40$^\circ$;][]{CorradiSchwarz1993,Cuesta1995}.

We note two caveats regarding the structural schematic view of NGC 6537 in Figure~\ref{fig:StructuralSchematic}. First, for simplicity, we have chosen not to include in this illustration a representation of the ionized nebular core itself, with its interior, bright-rimmed, ring or irregular spiral feature, and protruding, bubble-like structures. The JWST imaging of this ionized core region, which likely marks the UV-bathed and fast-wind-heated interior portions of the equatorial torus (Figure~\ref{fig:JWST_Bra_PAH_FeII_H2_RGB}, middle and bottom panels), is shown as an inset in the color overlay above the schematic diagram. Second, while the approximate line of sight toward this system of structural components \citep[corresponding to the inferred inclination of the nebular polar axis, $\sim$30--40$^\circ$;][]{CorradiSchwarz1993,Cuesta1995} is indicated in the diagram, the specific alignment of the polar lobe components with respect to each other and the molecular torus in this schematic (re)projected view of NGC 6537 is merely representative. The 3D orientation of these components remains subject to refinement via comprehensive, high-resolution optical and near-IR spectroscopic measurements of their (ionized gas) kinematics. 

Those caveats aside, the picture that emerges from the combined JWST/HST/ALMA view of NGC 6537 (represented schematically in Figure~\ref{fig:StructuralSchematic}) is that of a central star system that has formed a bipolar PN via a complex and (evidently) ongoing sequence of stellar mass loss events. The ejection of the molecular torus, whose onset began $\sim$10000 yr ago and is characterized by AGB-like expansion velocities, likely marked the termination of the AGB evolutionary stage of the relatively high-mass nebular progenitor. Beginning a few$\times$1000 yr after the onset of torus ejection, that slow, dense, equatorially-directed outflow was succeeded by, and perhaps helped collimate, the fast polar outflows. Furthermore, the point-symmetric shocks that are marked by the extended [Fe {\sc ii}] emission regions along the opposing sides of the polar lobe walls, as well as the sharp-rimmed structures within and around the ionized nebula core, serve as evidence that this fast outflow stage continues at present, and that the main direction of outflow has precessed or wandered considerably during the epoch of polar lobe formation.  

This scenario for the formation and evolution of NGC 6537 is essentially the same as that postulated for NGC 6302, on the basis of HST, ALMA, and JWST observing campaigns targeting the latter bipolar PN \citep[][]{Kastner2022,Balick2023,MoragaBaez2025,Matsuura2025}, as well as for the Ring and Southern Ring nebulae (NGC 6720 and NGC 3132), on the basis of JWST and mm-wave CO observations of these PNe \citep{Kastner2024,Kastner2025}. In each case, the sequence of slow, dense equatorial outflow followed by fast, collimated polar outflows --- which, as noted above (\S \ref{sec:COtorus}), appears to be a common characteristic of PNe with torus/lobe or torus/jet structures --- points to the influence of a binary companion to the central, mass-losing post-AGB star that has generated the PN. We return to the evidence in favor of such an interacting binary model for NGC 6537 below (\S \ref{sec:IRexcess}).

\subsection{Hydrodynamic modeling}\label{sec:hydro}

\begin{figure}[ht]
\begin{center}
\includegraphics[width=6.5in]{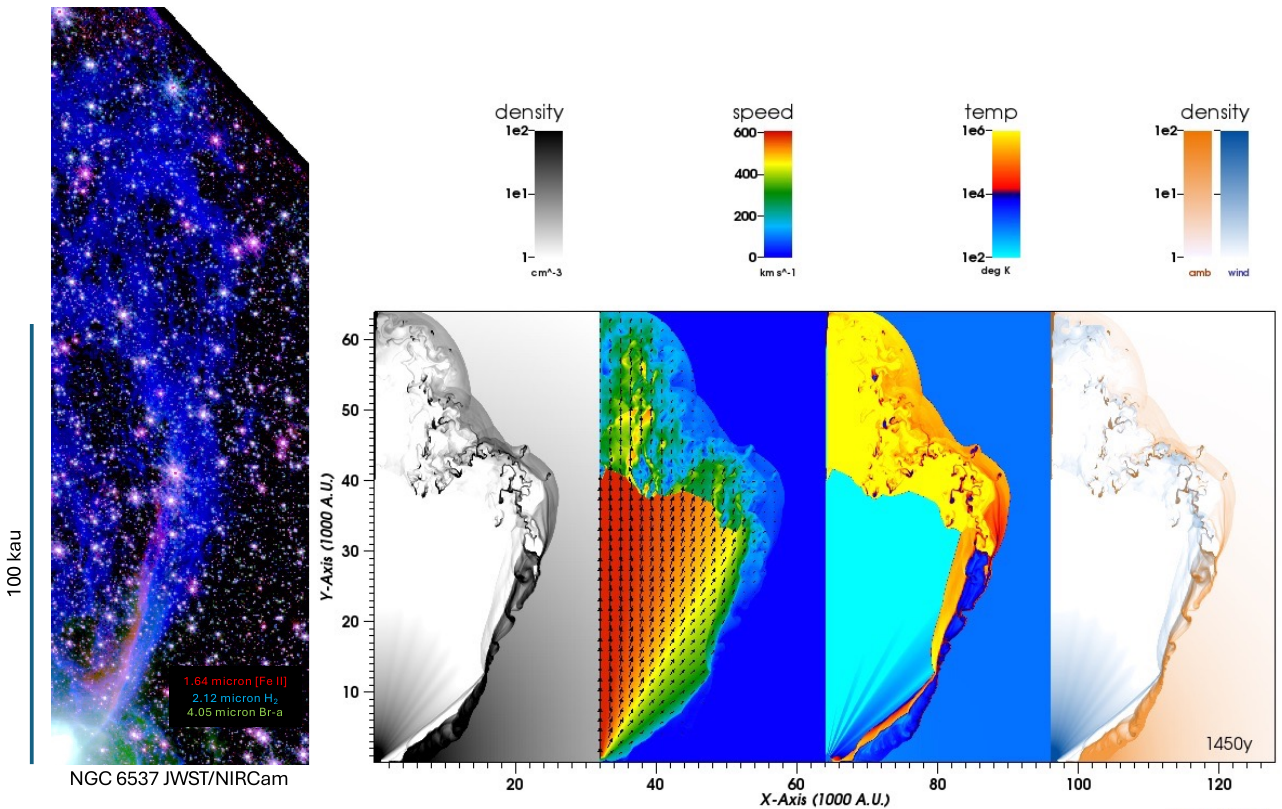}
\end{center}
\caption{Overlay of 4.05 $\mu$m Br$\alpha$ (green), 2.12 $\mu$m H$_2$ (blue), and 1.64 $\mu$m [Fe {\sc ii}] (red) images of half of the NE-directed polar lobe of NGC~6537, with the lobe ``half'' flipped along its polar axis (leftmost panel), compared with images illustrating various aspects of the simulation described in \S~\ref{sec:hydro}.}
\label{fig:JWSTvsBruceModel}
\end{figure}

To understand the physical mechanisms that might explain the basic structure of NGC 6537 illustrated in Figure~\ref{fig:StructuralSchematic}, we turn to the recent hydrodynamic  simulations of evolved bipolar nebula formation and growth presented in \citet[][]{Balick2019,Balick2020}. The reader is directed to those papers for details of the simulations. The basic model consists of fast, azimuthally tapered, post-AGB polar outflows (jets) steadily impinging on slower-moving AGB ejecta whose spherical density profile scales as $R^{-2}$. The model is agnostic as to the physical mechanism(s) responsible for the fast outflows and the pre-existing AGB outflow. 

The specific, simple, two-dimensional simulation presented in Figure~\ref{fig:JWSTvsBruceModel} is selected as representative of a ``family'' of simulations included in \citet[][]{Balick2019} that appears to match the overall radial and azimuthal appearance of the polar lobes of NGC 6537. The simulation is intended only to be illustrative of the basic physics describing the growth of the polar lobes. The basic parameters of the model include a pre-existing and stationary (``slow'') AGB wind with a radial density profile given by $n_{\mathrm{slow}}(r) = 5000 \, \mathrm{cm}^{-3} \times (2 \, \mathrm{kau}/r)^2$ and a fast wind whose axial density profile is given by $n_{\mathrm{fast}}(r) = 1000 \, \mathrm{cm}^{-3} \times (2 \, \mathrm{kau}/r)^2$ and axial speed $v_{\mathrm{fast}} = 600$ km s$^{-1}$.  Both the density and speed are tapered in the azimuthal direction, $\theta$, by a Gaussian whose $1/e$ width is 60$^\circ$. The diagnostic images in Figure~\ref{fig:JWSTvsBruceModel} represent the simulation after a duration of 1450 yr, about half the measured kinematic age of the lobes of NGC 6537. At this age, the forward speed of the leading edge of the lobe at $\theta = 0$ has slowed to 130 km s$^{-1}$, or ~25\% of the fast wind ejection speed, owing to the displacement of the denser slow wind downstream.  Features along the lobe edges also move far more slowly --- at speeds far less than the nearby fast winds whose momentum accelerated them --- and their speeds steadily decrease with radius.  Thus, in this simulation, the expansion ages of features along the lobe edges can significantly underestimate the time since the onset of the fast wind. Other details concerning the model methodology, as well as the early-time structure of the simulation presented here, are described in \citet[][see their \S 3.3]{Balick2022}.

A comparison of this representative hydrodynamic model with a NIRCam image overlay for half of the NE polar lobe of NGC 6537 is presented in Figure~\ref{fig:JWSTvsBruceModel}. Notwithstanding the difference in dynamical ages between data and model, it is apparent that the simulation reproduces many of the salient features of the JWST imaging. In particular, the simulated polar lobe opening angles closely match the image data, and the thin edges of the model lobes are closed, highly flocculent (turbulent), and likely perforated, as imaged in H$_2$. Moreover, the narrow high-temperature wind interaction region along the inner edge of the lobe walls --- which appears red-orange in the rendering of the simulation temperature map, indicative of a gas temperature in the range $\sim$$10^4$--$10^5$ K --- closely matches the position and width of the [Fe {\sc ii}]-emitting regions detected by NIRCam. 

The comparison in Figure~\ref{fig:JWSTvsBruceModel} hence supports the basic nebular formation model described schematically in \S \ref{sec:schematic} and Figure~\ref{fig:StructuralSchematic}.  We caution, however, that the methodology underlying the simulation illustrated in Figure~\ref{fig:JWSTvsBruceModel} was not designed to model the physical conditions resulting from a highly directed jet whose outflow direction is likely directionally variable, and is presently (and perhaps persistently) misaligned with the axis of symmetry of the polar lobes, such as appears to be responsible for the [Fe {\sc ii}] emission along opposing lobe walls imaged in NGC 6537 \citep[as well as in NGC 6302;][]{Kastner2022}. Additional hydrodynamical modeling that is specifically tailored to such a ``wandering'' polar outflow configuration is clearly warranted.

\subsection{IR excess at the central star: smoking gun of binarity?}\label{sec:IRexcess}

\begin{figure}[ht]
\begin{center}
\includegraphics[width=3.5in]{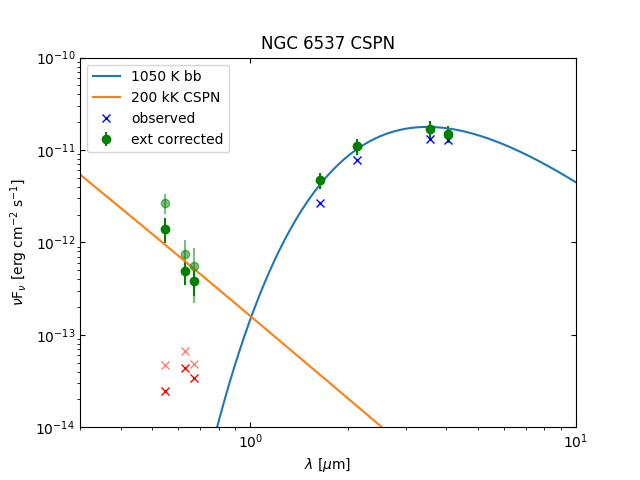}
\end{center}
\caption{Spectral energy distribution (SED) of the CSPN of NGC 6537 from 0.55 $\mu$m to 4.05 $\mu$m, as extracted from (archival) HST/WFPC2 and (our new) JWST/NIRCam narrow-band images (red and blue x's, respectively). Dark and light red x's indicate our photometry and photometry reported in \citet{Matsuura2005N6537}, respectively. The extinction-corrected photometry \citep[assuming $A_V = 3.7$;][]{Matsuura2005N6537} is displayed as green symbols, with dark and light symbols for our results and those reported in \citet{Matsuura2005N6537}, respectively. The SED is overlaid with blackbodies representing the hot central star photosphere (orange curve) and hot dust (blue curve) that have been (arbitrarily) scaled to the extinction-corrected photometry, for reference.}
\label{fig:NGC6537CSPNSED}
\end{figure}

To confirm and further investigate the nature of the IR excess of the central star that is evident in 
Figure~\ref{fig:JWSTzoom2} as well as Figure~\ref{fig:JWST_Bra_PAH_FeII_H2_RGB}, we extracted photometry for the star from the new JWST/NIRCam and archival HST/WFPC2 images. 
As the PSFs of the HST and JWST images are all $<0.3''$ (FWHM) and are reasonably well matched, we used an aperture of radius of 0.15$''$ and an annulus with inner and outer radii of 0.15$''$ and 0.3$''$ to measure the central star and background fluxes, respectively, within the four JWST/NIRCam images as well as the HST/WFPC2 F547M, F631N, and F673N filter images. The latter set of (HST) images are those for which the bandwidths are relatively wide and line emission is relatively weak and, hence, the central star is clearly detected \citep[see][]{Matsuura2005N6537}. 
The resulting central star fluxes, corrected for reddening assuming $A_V = 3.7$ \citep[][]{Matsuura2005N6537}, are presented in Fig.~\ref{fig:NGC6537CSPNSED}, where we also include the extinction-corrected flux measurements reported in \citet[][]{Matsuura2005N6537}. The errors in our flux measurements are dominated by uncertainties in the determination of the bright nebular background, and different background subtraction strategies likely account for the slight discrepancies between our photometry and that determined by \citet[][]{Matsuura2005N6537}.

The resulting 0.55 $\mu$m to 4.05 $\mu$m spectral energy distribution (SED) of the central star of NGC 6537 in Figure~\ref{fig:NGC6537CSPNSED} clearly demonstrates the presence of a significant IR excess, extending across all four JWST filters. The photospheric temperature of the central star, which is poorly constrained by the HST/WFPC2 data \citep[see][their Table 3]{Matsuura2005N6537}, is represented by a 200 kK blackbody model in the Figure.  The IR excess can then be well matched by a model blackbody that peaks at a wavelength of roughly 3.5 $\mu$m, corresponding to a blackbody temperature of $\sim$1000 K, i.e., approaching dust sublimation conditions. This suggests that the excess IR radiation from the central star's vicinity, if due to thermal emission from dust, emanates from grains heated nearly to the point of sublimation by the luminous central star's intense radiation field. From simple thermal equilibrium considerations, the emitting dust would lie at a characteristic distance of $\sim$7 au from the star, assuming a stellar luminosity of 8500 $L_\odot$ \citep{Casassus2000}.

The bipolar PN NGC 6537 thus can be added to the growing list of PNe, most of them molecule-rich, for which the central star exhibits an IR excess due to dust in an unresolved or only marginally resolved circumstellar structure --- beginning with the detection by the {\it Spitzer Space Telescope} of an IR excess at the central star of the Helix Nebula \citep[NGC 7923;][]{Su2007} and, in the era of JWST, continuing with detections of IR excesses at the central stars of the Ring and Southern Ring nebulae \citep[NGC 6720, NGC 3132;][]{DeMarco2022,Sahai2023,Sahai2025} as well as NGC 6537's close analog, NGC 6302 \citep[Wesson et al.\ 2025, in prep.; see also][]{Matsuura2025}. In all four cases just mentioned, detections of the PN central star IR excesses required JWST's unique combination of (infrared) wavelength coverage and subarcsecond imaging resolution. However, of the four, only the stellar photosphere of NGC 6302 has escaped direct detection by both HST and JWST, as a consequence of the near edge-on orientation of its extensive, optically thick torus \citep[][]{Kastner2022,Matsuura2025}.  In contrast to NGC 6302, and in spite of the many similarities between this nebula and NGC 6537, the photospheric emission of the central star of NGC 6537 is directly detectable and measureable (Fig.~\ref{fig:NGC6537CSPNSED}) thanks to the intermediate inclination of its dusty, expanding molecular torus (Figure~\ref{fig:StructuralSchematic}).

Most investigations of PN central star IR excesses have concluded that the circumstellar dust responsible for the excess resides in a long-lived, orbiting disk, most likely in a circumbinary configuration, although the mechanisms that might generate a circumbinary dust disk in evolved star systems remain subject to vigorous debate \citep[see, e.g.,][and references therein]{Clayton2014,DeMarco2022,Sahai2025}. By analogy with the aforementioned previous cases of IR excesses at the central stars of PNe, the most likely explanation for the presence of the compact (unresolved) hot dust source detected within NGC 6537 in our JWST imaging is that the central star resides in a binary system that is orbited by a circumbinary disk. 

While such a (Keplerian disk) configuration would appear to stand in stark contrast to the expanding molecular torus we have detected in $^{13}$CO emission (\S\ref{sec:COtorus}) in terms of both size scale and kinematics, these two circumstellar structures may be complementary, rather than contradictory: both point to the influence of a binary companion on the mass loss history of the progenitor star of NGC 6537. That is, the same (putative) companion responsible for formation of a dusty circumbinary disk also may have focused the primary's pre-PN (AGB epoch) mass loss along the equatorial plane of the system \citep[e.g.,][]{MastrodemosMorris1999}, 
thereby generating the dusty, molecular torus and the fast, collimated polar outflows that followed torus formation. The fact that the polar outflows in NGC 6537 appear to change direction over time (\S \ref{sec:schematic}) is, again, very similar to the variable-direction, point-symmetric outflows observed in NGC 6302 \citep{Balick2023} as well as in many other molecule-rich, bipolar and multipolar PNe that are both dynamically younger \citep[e.g., NGC 7027;][]{MoragaBaez2023} and older \citep[e.g., NGC 6072;][]{Kwok2010}.  Such wandering or precessing collimated outflows or jets have been proposed as the natural outcomes of interacting binary systems involving mass-losing AGB stars, either in detached configurations in which the secondary captures a fraction of the AGB primary's ejected mass and forms an accretion disk \citep[e.g.,][and references therein]{MastrodemosMorris1998,Sahai2016}, or in closer configurations that lead to the companion's plunging into the AGB envelope \citep[e.g.,][and references therein]{Soker2019}.

\section{Summary and Conclusions}

We have presented new James Webb Space Telescope near-infrared (NIRCam) Br$\alpha$, H$_2$, [Fe {\sc ii}], and PAH imaging of the dusty, molecule-rich, pinched-waist, high-excitation bipolar planetary nebula (PN) NGC 6537 (the Red Spider Nebula). The combination of JWST/NIRCam 4.05 $\mu$m Br$\alpha$ and archival ALMA 3mm continuum images provides an extinction-penetrating view of the ionized nebular core, highlighting its complex structure. This ionized core region is shown to consist of a highly asymmetric ring or irregular spiral of radius $\sim$5$''$ ($\sim$0.04~pc) with surrounding bubble or loop structures that is superimposed on the inner rims of the nebula's inclined bipolar lobes. Immediately surrounding the ionized core region imaged by JWST and HST is a point-symmetric, expanding, equatorial molecular torus of projected radius $\sim$0.08 pc and projected outflow velocity $\sim$10 km s$^{-1}$ that is traced by ALMA $^{13}$CO(1--0) emission-line mapping. Assuming the polar axis of the nebula is inclined by 38$^\circ$ with respect to the plane of the sky \citep{CorradiSchwarz1993}, the deprojected radius ($\sim$0.13 pc) and expansion velocity ($\sim$13 km s$^{-1}$) of the molecular torus yield a torus dynamical age of $\sim$10000 yr.  The Br$\alpha$ and 3 mm continuum images (as well as archival HST H$\alpha$ imaging) evidently reveal photoionized and/or shock-ionized gas in the interior regions of this dense equatorial torus. 

JWST/NIRCam imaging of 2.12 $\mu$m H$_2$ emission traces the full extent of the polar lobes, demonstrating that they are in fact closed bubble-like structures with projected extents of $\sim$100$''$ ($\sim$0.8~pc). Given the deprojected lobe outflow velocities ($\sim$300~km~s$^{-1}$) and polar axis inclination inferred from optical spectroscopy \citep{CorradiSchwarz1993}, their deprojected $\sim$0.9 pc lengths indicate that the lobes have formed over dynamical timescales of $\sim$3700 yr.  The H$_2$-emitting polar lobe walls lie just outside the ionized lobe rims; these H$_2$-emitting outer lobe regions are highly structured, displaying delicate filamentary features indicative of local condensations likely resulting from instabilities in the nebula's significant molecular gas component. The PAH emission closely traces this 2.12 $\mu$m H$_2$ emission but, unlike the H$_2$ emission, the PAH emission displays a rapid drop in surface brightness with distance from the central star, indicating that the PAH and H$_2$ emission are excited by central star UV and slow ($\sim$10--20~km~s$^{-1}$) shocks, respectively. Ionized gas found in the foreground and background regions of the inner lobes is traced by the JWST/NIRCam Br$\alpha$ image and an archival HST/WFPC2 H$\alpha$ image, respectively.  

In striking contrast, the 1.64 $\mu$m [Fe~{\sc ii}] image displays an extended S-shaped morphology, with the [Fe~{\sc ii}] emission confined to narrow arcs along the inner eastern and western rims of the NE- and SW-directed lobes. This [Fe~{\sc ii}] emission morphology very closely resembles that seen in the similarly extreme (molecule-rich, high-excitation) bipolar PN NGC 6302. As in that case, the [Fe~{\sc ii}] in NGC 6537 likely traces fast ($\sim$100~km~s$^{-1}$) shocks from collisions of an active, collimated wind emanating from the vicinity of NGC 6537's central star with slower-moving material forming the polar lobe rims; such a fast wind is indicative of the presence of a close (interacting) companion star. Deep Chandra/HRC-I imaging of NGC 6537 yields no detection of X-rays from these shocks despite the high speeds of the collimated outflows and the presence of the [Fe~{\sc ii}] emission, indicating that the diffuse X-ray luminosity of the shocked regions of NGC 6537 is at least a factor $\sim$5 smaller than the X-ray luminosities of younger, X-ray-detected PNe that are rich in molecular gas \citep[i.e., BD+30 3639, NGC 7027;][]{Kastner:2000rr,MontezKastner2018}.

The NIRCam imaging demonstrates that the central star has a near-IR excess, indicating the presence of hot ($\sim$1000~K) circumstellar dust. By analogy with other PNe where such a central star near-IR excess is detected, the dust at NGC 6537's central star(s) may reside in an orbiting disk, most likely in a circumbinary configuration. The putative companion to the central star may also be responsible for the fast, collimated outflows that have inflated the polar lobe bubbles traced by near-IR H$_2$ emission and are presently generating the S-shaped structure seen in [Fe~{\sc ii}] emission. The nondetection of X-rays from the vicinity of the central star by Chandra/HRC-I suggests that any such close companion, if a main-sequence star, is less massive than the Sun.

We present a basic model for the structure and formation of NGC 6537. This model consists of a slow, dense equatorial outflow, ejected $\sim$10000 yr ago, that was impinged upon by more recent (dynamical age $\sim$3700 yr) fast, collimated outflows that are likely both episodic and precessing. This general scenario is supported by a hydrodynamic simulation of nebular growth, incorporating these same basic structural and kinematic components, that appears to well reproduce many of the salient features observed in the JWST and ALMA imaging. We assert that this model for the formation and evolution of NGC 6537 likely applies to most, if not all, molecule-rich, bipolar planetary nebulae that are descended from relatively massive progenitors. Such a sequence of slow, dense equatorial outflow followed by fast, collimated polar outflows again points to the profound influence of an interacting binary companion to the central, mass-losing primary star on the primary's late evolution and its mass loss rate and geometry, with this influence beginning during the AGB stage and continuing through the present epoch. 

\vspace{.5in}

{\it Acknowledgements.} The authors thank the anonymous referee for valuable comments that improved this paper. Support for this research was provided by Space Telescope Science Institute grant JWST--GO--04571.001--A to RIT, and by the NSF through the National Radio Astronomy (NRAO) Student Observing Support program via award SOSPADA-009 to RIT. Research by J.K. and P.M.B. on the molecular content of planetary nebulae is further supported by NSF grant AST--2206033 to RIT.  M.S.-G. acknowledges the financial support of I+D+i project PID2023-146056NB-C21, funded by the Spanish MICIU/AEI//10.13039/501100011033 and ERDF/UE. This work is based on observations made with the NASA/ESA/CSA James Webb Space Telescope. The data were obtained from the Mikulski Archive for Space Telescopes at the Space Telescope Science Institute, which is operated by the Association of Universities for Research in Astronomy, Inc., under NASA contract NAS 5-03127 for JWST. These observations are associated with program \# 4571. The JWST and HST data presented in this article were obtained from the Mikulski Archive for Space Telescopes (MAST) at the Space Telescope Science Institute; the specific JWST and HST observations analyzed can be accessed via \dataset[doi: 10.17909/mvvm-p397]{https://doi.org/10.17909/mvvm-p397} and \dataset[doi: 10.17909/z0bw-7m72]{https://doi.org/10.17909/z0bw-7m72}, respectively. This paper employs a list of Chandra datasets, obtained by the Chandra X-ray Observatory, contained in the Chandra Data Collection \dataset[doi: 10.25574/cdc.462]{https://doi.org/10.25574/cdc.462}. This paper makes use of ALMA data ADS/JAO.ALMA\#2018.1.00424.S. ALMA is a partnership of ESO (representing its member states), NSF (USA) and NINS (Japan), together with NRC (Canada), NSTC and ASIAA (Taiwan), and KASI (Republic of Korea), in cooperation with the Republic of Chile. The Joint ALMA Observatory is operated by ESO, AUI/NRAO and NAOJ. The National Radio Astronomy Observatory is a facility of the National Science Foundation operated under cooperative agreement by Associated Universities, Inc.

\facility{JWST, HST, ALMA, Chandra}




\end{document}